\documentclass[a4paper,prl,twocolumn,longbibliography,sort&compress]{revtex4-1}
\usepackage{graphicx}  
\usepackage{bm}        
\usepackage{amsmath}   
\usepackage{amssymb}   
\usepackage{amsmath}
\usepackage{siunitx} 

\usepackage{color}
\newcommand{\ml}[1]{#1}

\date{\today}
\begin{document}
\title{Mxdrfile: read and write Gromacs trajectories with Matlab}
\author{Jon Kapla}
\email[]{jon.kapla@mmk.su.se}
\affiliation{Department of Materials and Environmental Chemistry,
  Stockholm University, SE-106\,91 Stockholm, Sweden.}
\altaffiliation{Current address: Department of Medicinal Chemistry,
  Uppsala University, SE-751\,23 Uppsala, Sweden}
\author{Martin Lind\'en}
\affiliation{Department of Cell and Molecular Biology, Uppsala
  University, 751 24 Uppsala, Sweden.}
\altaffiliation{Current address: Scania CV AB, Södertälje, Sweden}

\begin{abstract}
Progress in hardware, algorithms, and force fields are pushing the
scope of molecular dynamics (MD) simulations towards the length- and
time scales of complex biochemical processes. This creates a need for
developing advanced analysis methods tailored to the specific
questions at hand. We present mxdrfile, a set of fast routines for
manipulating the binary xtc and trr trajectory files formats of
Gromacs, one of the most commonly used MD codes, with Matlab, a
\ml{powerful and versatile language for scientific computing.  The 
unique ability to both read and write binary trajectory files makes it
possible to leverage the broad capabilities of Matlab to speed up and
simplify the development of complex analysis and visualization
methods.}
We illustrate these possibilities by \ml{implementing an alignment
method for buckled surfaces, and use it to} briefly dissect the
curvature-dependent composition of a buckled lipid bilayer. The
mxdrfile package, including the buckling example, is available as open
source at \url{http://kaplajon.github.io/mxdrfile/}.
  
\end{abstract}

\maketitle

\section{Introduction}
Molecular dynamics (MD) simulations are useful complements to
experimental studies of many biomolecular processes, because they
generate information about all classical degrees of freedom of the
system with atomic resolution by solving Newton's equations of motion.

The length- and time-scales that are accessible to atomistic MD
simulations are increasing rapidly, due to increasing computer power
and improved algorithmic performance, which means that it becomes
feasible to simulate increasingly complex systems and processes
\cite{pronk2013}. Even larger and longer simulations are possible by
using coarse-graining, where detailed structural features are
systematically removed from the models in order to increase
computational speed \cite{reynwar2007,marrink2007}. Simulations of
complex processes often require customized analysis and visualization
methods that are most easily developed using high-level languages for
general-purpose scientific computing. This, in turn, requires efficient
methods to read and write the simulated MD trajectories in such
environments which is not a trivial task, as trajectory files often
use non-intuitive formats to minimize file size.

Here, we present the mxdrfile toolbox, which can read and write
binary trajectories from one of the most popular and powerful MD
codes, Gromacs \cite{pronk2013}, with Matlab, a powerful and widely
used language for general-purpose scientific computing. Our routines
\ml{utilize} the loadlibrary functionality in Matlab to access the
binary files directly \ml{through} the Gromacs xdr library, which was
developed precisely to allow easy interfacing with external analysis
tools \cite{xdrfile}. This offers an efficient and robust
implementation, and extends the functionality of previous Gromacs
parsers for Matlab which are read-only
\cite{arthur,ribarics,dien2014}.

\section{Basic usage}
Reading or writing Gromacs trajectories (xtc or trr files) are done in
three steps: opening a binary file for reading or writing, stepping
through the trajectory to read or write each frame, and finally
closing the file. Using the wrapper commands of mxdrfile, loading and
parsing a trajectory requires about 10 lines of code: {\small
\begin{verbatim}
 loadmxdrfile; % Load the xdrfile library 
 % open files for reading and writing
 [~,rTraj]=inittraj('test.xtc','r');
 [~,wTraj]=inittraj('testwrite.xtc','w');
 % read first frame
 [rstatus,traj]=read_xtc(rTraj);
 % parse trajectory
 while(not(rstatus))
    % Do something with the coordinates
    traj.x.value=traj.x.value*0.8;
    % Write newcoords to a new xtc file
    wstatus=write_xtc(wTraj, traj);
    % read next frame
    [rstatus,traj]=read_xtc(rTraj);
 end
% close files
[status,rTraj]=closetraj(rTraj);
[status,wTraj]=closetraj(wTraj);
\end{verbatim}}

\begin{figure}
  \includegraphics[width=80mm]{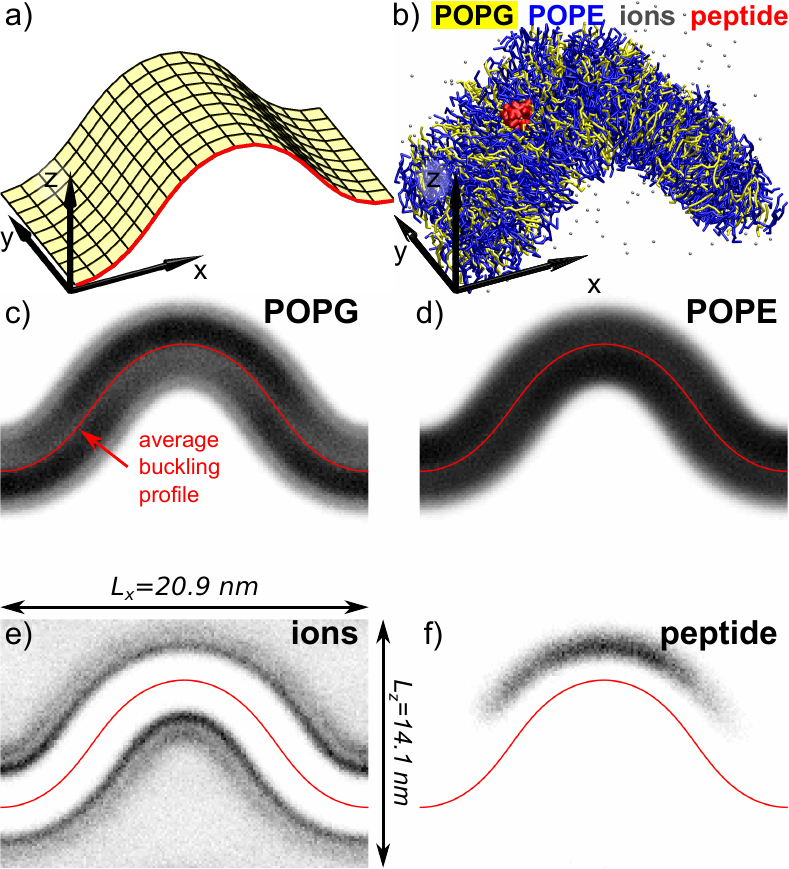}
  \caption{\label{fig} Component distributions in a coarse-grained
    buckled POPE/POPG bilayer interacting with a single Magainin
    peptide \cite{gomez2016}. a) Theoretical buckled surface used for
    alignment. b) Simulation snapshot. c-f) Mass density plots in the
    $xz$-plane of vrious components. Molecular graphics generated with
    VMD \cite{vmd96}.}
\end{figure}
\ml{For read-only parsing, an @mxdrfile class provides even more compact
access to both trajectory types, as illustrated by the following
minimal example:}
{\small
\begin{verbatim}
 loadmxdrfile; % Load the xdrfile library
 % open file and read the first frame
 trj=mxdrfile('test.xtc');
 z=[];
 % Loop over the trajectory frames
 while(~trj.status)
     z(end+1)=mean(trj.x.value(3,:));
     trj.read(); % move to next frame
 end
 clear trj % also closes trj file
\end{verbatim}}
\section{Case study: membrane buckling}
Next, we consider an application that makes use of Matlab's numerical
capabilities, by studying the curvature-dependent composition of a
lipid bilayer using simulated buckling.  Buckling a lipid bilayer by
compressing the projected area of a rectangular patch creates a shape
that closely follows the classical Euler buckling profile
\cite{noguchi2011,hu2013}, and thus presents a range of positive and
negative curvatures that makes it an excellent model system for
simulating membrane curvature sensing \cite{gomez2016} (see
Fig.~\ref{fig}a,b). The shape is maintained by simulating an ensemble
where the projected area is kept fixed (but the box height may be used
to maintain constant pressure).

However, to extract useful information, one must remove random drift
of the membrane profile from the simulation in order to analyse
motions relative to the buckled membrane profile.  This presents a
challenging alignment problem.  Since the lipids move randomly within
the membrane plane, simply minimizing the RMSD to some reference
configuration will not work well. An alternative approach is to use
the theoretical Euler buckling profile (Fig.~\ref{fig}a) as a
template, and align by minimizing the total square distance between
the buckling profile and the beads at the end of the lipid tails,
which are situated near the membrane midplane. The degree of buckling
of the template shape should also be flexible to accomodate small area
fluctuations of the membrane.

The Euler shape depends on a dimensionless compression factor
$\gamma=(L-L_x)/L$, where $L$ is the arclength along the buckled
direction, and $L_x$ is the corresponding projected length. Hence, if
we parameterize the buckled profile as a function of $\gamma$ for a
reference size, the general case can be obtained by shifting and
scaling. Following ref.~\cite{gomez2016}, we choose $L_x=1$ as
reference, and write the general buckling profile in the $x,z$ plane
as a parametric curve
\begin{equation}
  x=L_x\big(s+\xi(s,\gamma)\big),\quad
  z=L_x\zeta(s,\gamma),
\end{equation}
where $s$ is an arclength-like parameter that we will take to be
normalized to have period 1. If the beads to align have coordinates
$\{x_i,z_i\}$, $i=1,2,\ldots$, the alignment problem consists of
minimizing the total square distance
\begin{equation}
  \sum_i \Big(x_0+L_x\big(s+\xi(s_i,\gamma)\big)-x_i\Big)^2
  +\Big(z_0+L_x\big(\zeta(s_i,\gamma)\big)-z_i\Big)^2
\end{equation}
w.r.t.~the shape parameter $\gamma$, the projected positions $s_i$,
and the overall displacements $x_0,z_0$.


We used mxdrfile to solve this alignment problem in Matlab.  An
efficient numerical solution of this non-linear least-squares problem
requires good minimization routines (which Matlab has) as well as fast
evaluation of $\xi,\zeta$ and their partial derivatives
w.r.t~$s,\gamma$. The Euler buckling problem has a \ml{partly implicit
  analytical solution in terms of elliptic functions
  \cite{noguchi2011}. We chose a brute-force numerical approach
  instead, and used}
Matlab's built-in boundary value solver to compute buckling profiles
for the reference case $L_x=1$, from which we expanded $\xi,\zeta$ in
truncated Fourier series that for symmetry reasons take the forms
\begin{align}
  \xi(s,\gamma)=&  \sum_{n=1}^N a_n^{(x)}(\gamma)\sin(4\pi ns),\\
  \zeta(s,\gamma)=&a_0^{(z)}(\gamma)
  +\sum_{n=1}^N a_n^{(z)}(\gamma)\cos(2\pi(2n-1)s).
\end{align}
Only a few terms are needed, and partial derivatives w.r.t.~$s$ are now
easily constructed. For the coefficients $a_j^{(x,z)}(\gamma)$, we
used Matlab's piece-wise polynomial interpolation, which can be
exactly differentiated w.r.t.~$\gamma$. 

We applied the above analysis to a 15 \si{\micro\second} Martini
\cite{marrink2007,dejong2013} simulation of the antibacterial
amphipathic peptide magainin interacting with a buckled two-component
lipid bilayer (70:30 POPE:POPG, see Ref.~\cite{gomez2016} for
details). Fig.~\ref{fig}c shows transverse density profiles of the two
lipid species, the counter ions, and the peptide, all of which display
non-trivial curvature-dependent distributions. An animation showing
part of the trajectory before and after alignment is shown in
supplementary movie S1.

\section{Conclusions}
We provide a set of \ml{routines} to enable read and write access to
Gromacs binary xtc and trr trajectory file formats in Matlab. The code
was showcased with an application to solve the alignment problems
related to Euler buckling of lipid bilayers. We believe that the
ability to parse and manipulate Gromacs trajectories will be similarly
useful in tackling future complex simulation problems, utilizing the
full potentials of Gromacs and Matlab.

The mxdrfile package, including the buckling
example, is available as open source at
\url{http://kaplajon.github.io/mxdrfile/}.


\paragraph{Acknowledgements}
Finanical support from the Carl Trygger Foundation (J.K. through
Arnold Maliniak), the Wenner-Gren Foundations (M.L.) and the Swedish
Foundation for Strategic Research via the Center for Biomembrane
Research (M.L.) are gratefully acknowledged.

%


%

\end{document}